\documentclass[12pt]{article}
\usepackage{threeparttable}
\usepackage{amsmath}
\usepackage{amssymb}
\usepackage{natbib}

\begin{document}
\bibliographystyle{plainnat}
\setcitestyle{numbers,square}

\title{\normalsize{RADIAL PULSATIONS OF NEUTRON STARS: COMPUTING \\
                   ALTERNATIVE POLYTROPIC MODELS REGARDING \\
                   DENSITY AND ADIABATIC INDEX}
      }
\author{Vassilis Geroyannis$^1$, Georgios Kleftogiannis$^2$ \\
        $^{1,2}$Department of Physics, University of Patras, Greece \\
        $^1$vgeroyan@upatras.gr, $^2$gkleftis@upatras.gr}
\maketitle

\newcommand{\be}{\begin{equation}}
\newcommand{\ee}{\end{equation}}

\begin{abstract}
We revisit the problem of radial pulsations of neutron stars by computing four  general-relativistic polytropic models, in which ``density'' and ``adiabatic index'' are involved with their discrete meanings: (i) ``rest-mass density'' or (ii) ``mass-energy density'' regarding the density, and (i) ``constant'' or (ii) ``variable'' regarding the adiabatic index. Considering the resulting four discrete combinations, we construct corresponding models and compute for each model the frequencies of the lowest three radial modes. Comparisons with previous results are made. The deviations of respective frequencies of the resolved models seem to exhibit a systematic behavior, an issue discussed here in detail. 
\end{abstract}

\section{Introduction}
The study of radial pulsations of relativistic stars is an interesting issue. By computing the radial modes of a stellar model, we obtain significant information about the stability of the model. Radial modes were first investigated in \cite{CH64}. Since then, they have been studied by several authors for various stellar models, focusing on neutron stars \citep{METH66,CHN77,GLLI83,VACHN92,KORU01}, protoneutron stars \citep{GOHA97}, and strange stars \citep{VACHN92,GOZD99}, obeying several ``equations of state'' (EOS, EOSs). 

An EOS often used is the polytropic EOS, since its analytical form makes easier the numerical computations. In the bibliography, we find two discrete interpretations of the polytropic EOS. In particular, in the first interpretation the mass-energy density, $E$, enters the polytropic EOS (see e.g. \citep{TOOP64}). In the second interpretation, on the other hand, the rest-mass density, $\rho$, is that entering the polytropic EOS (see e.g. \citep{TOOP65}). Both interpretations have been adopted by several authors treating radial pulsations (see e.g. \citep{KORU01} for the first interpretation, and \citep{HARTLEVIII} for the second interpretation), as well as of nonradial oscillations (see e.g. \citep{CAMPO67}). 

On the other hand, the adiabatic index $\Gamma$, which enters the equations governing the radial pulsations, can be defined in several ways dependent on the physical conditions assumed. The difference between the two of the definitions, that we give emphasis on, concerns the variability of $\Gamma$. In this view, there are authors assuming a variable $\Gamma$ (see e.g. \citep{GLLI83} and \citep{KORU01}), as well as authors assuming a constant $\Gamma$ (see e.g. \citep{HARTLEVIII}).

Since it is a common practice for researchers to employ the polytropic EOS for testing new codes, it may happen that numerical results obtained by a particular interpretation of the polytropic EOS and the adiabatic index be in discrepancy with those compared with, not due to some errors of a particular code, but rather due to different interpretations of the polytropic EOS and the adiabatic index.   
The aim of this investigation is to highlight the discrete interpretations of the polytropic EOS and the adiabatic index $\Gamma$, occuring in the bibliography, and to compute respective numerical results. We undertake such a task by combining the four assumptions made for the polytropic EOS and the adiabatic index $\Gamma$, and resolving the four discrete general-relativistic polytropic models constructed this manner.

\section{Theoretical Background}
Unless stated otherwise, all physical quantities are expressed in ``gravitational units", abbreviated ``gu'', in which the gravitational constant, $G$, and the speed of light, $c$, are equal to unity. To facilitate comparisons, we also give some significant quantities in ``polytropic units related to the gravitational units'', abbreviated ``pu''. A discussion on converting the well-known ``cgs units'', abbreviated ``cgs'', to gravitational units, and the latter to their related polytropic units can be found in \citep{GESF} (Sec.~1.2).

\subsection{The polytropic EOS}
\label{polEOS}
In the polytropic models, the pressure $P$ obeys the polytropic EOS    
\be\label{17}
P = K \, \rho^{1 + \left( 1/n \right)},
\ee
where $n$ is the well-known polytropic index, 
\be\label{18}
n=\frac{1}{\Gamma-1},
\ee
hence, when $\Gamma$ is to be  interpreted as constant,
\be\label{18plus} 
\Gamma = 1 + \frac{1}{n},
\ee
and the integration constant $K$ is the polytropic constant. 

Tooper (\citep{TOOP65}, Sec.~II) suggests that, due to the relativistic first law of thermodynamics (cf. \citep{TOOP65}, Eq.~(2)) 
\be\label{21}
\frac{d E}{E + P} = \frac{d\rho}{\rho},
\ee
the relation (cf. \citep{TOOP65}, Eq.~(4))  
\be\label{22}
\frac{dP}{P} = \Gamma \, \frac{d E}{E + P}
\ee
has to be satisfied.
This relation leads to the conclusion that the mass-energy density must be connected with the pressure via an equation of the form
\be\label{23}
E = C \, P^{1 / \Gamma} + \frac{P}{\Gamma - 1},
\ee
where $C$ is a constant. If we put $C = K^{-1 / \Gamma}$, then we find (cf. \citep{TOOP65}, Eq.~(5b)) 
\be\label{24}
E = \rho + n \, P,
\ee
with the polytropic index $n$ given by Eq.~(\ref{18}). It is worth clarifying here that the rest-mass density $\rho$ is that part of the mass density which satisfies a continuity equation and thus is conserved throughout the motion.  

Concerning the two forms of the polytropic EOS, i.e. Eq.~(\ref{17}) on the one side, and the EOS (cf. \citep{TOOP65}, Eq.~(7a)) 
\be\label{19}
P = K \, E^{1+\left( 1/n \right)}
\ee
on the other side, Tooper (\citep{TOOP65}, Sec.~II) remarks that Eq.~\eqref{19} permits the sound velocity to become greater than $c$ for all $n$, while Eq.~\eqref{17} gives a sound velocity less than $c$ for $n \ge 1$.

\subsection{The nonrotating model}
The Schwarzschild metric of a nonrotating spherical object, expressed in spherical coordinates $(r,\,\theta,\,\phi)$, is given by (cf. \citep{H67}, (Eq.~(25))
\be\label{1}
ds^2=-e^{\nu}dt^2+e^{\lambda}dr^2+r^2(d\theta^2+\sin^2\theta d\phi^2),
\ee
where $\nu$ and $\lambda$ are metric functions of $r$. For $\nu \ll 1$, the Newtonian gravitational potential, $\phi$, and the metric function $\nu$, expressed in cgs units, are related via the equation $\phi = c^2 \nu / 2$. It is therefore convenient to define an equivalent of the gravitational potential in general relativity as
\be\label{2}
\Phi  =\frac{\nu}{2}.
\ee
The function $\lambda$ is given by (cf. \citep{H67}, (Eq.~(27))
\be\label{3}
e^\lambda = \left( 1 - \frac{2m}{r} \right)^{-1},
\ee
where $m=m(r)$ is the mass contained within a sphere of radius $r$. 

Furthermore, a neutron star obeys the three equations of the general-relativistic hydrostatics:
\begin{enumerate}
\item The equation of hydrostatic equilibrium (cf. \citep{H67}, Eq.~(28)),
\be\label{5}
\frac{dP}{dr} = - \, \frac{(E+P)(m  +4 \pi \, r^3 \, P)}{r(r  -2m)}.
\ee

\item The equation of the mass-energy (cf. \citep{H67}, Eq.~(29a)), 
\be\label{6}
\frac{dm}{dr}  =4 \pi \, r^2 \, E.
\ee

\item The equation of the gravitational potential (cf. \citep{H67}, Eq.~(29b)), 
\be\label{7}
\frac{d\Phi}{dr} = - \, \frac{1}{E+P} \, \frac{dP}{dr}.
\ee
\end{enumerate}
Eqs.~\eqref{5} and \eqref{6} are the ``Oppenheimer-Volkof (OV) equations". Their solution must obey the initial conditions (cf. \citep{HARTLEII}, discussion following Eq.~(3b))
\be\label{9}
P(r=0) = \mathfrak{P}(E_c),
\ee
\be\label{8}
m(r=0) = 0,
\ee
where $E_c$ is the central mass-energy density. For the pressure $P$ we assume the relation   
\be\label{4}
P= \mathfrak{P}(E),
\ee
where by the symbol $\mathfrak{P}$ we emphasize on the fact that there is a particular functional relation assigned to the pressure $P$ with respect to the mass-energy density $E$, i.e. a particular EOS.

It is convenient to normalize $\Phi$ in the same manner as the Newtonian gravitational potential, that is,
\be\label{10}
\Phi(\infty) = 0.
\ee
This condition becomes at the boundary of the star (cf. \citep{GK08}, Eq.~(13))
\be\label{11}
\Phi(R) = \frac{1}{2} \left( 1- \frac{2M}{R} \right),
\ee
where $M$ is the total mass-energy of the star and $R$ is its radius.

\subsection{Equations governing the radial pulsations}
The general relativistic equations governing infinitesimal radial pulsations for a gas sphere were first derived in \citep{CH64}. Similar equations as a result of the slow rotation of a relativistic star have been derived in \citep{HARTLEVI}. Since then, the equations governing the radial oscillations were rewritten in various forms, some of them being suitable for numerical computations (see e.g. \citep{BTM66,CHN77,GLLI83,GOHA97}). In this study, we follow \citep{KORU01} (Sec.~2.1, and references therein) and write the relation (\citep{KORU01}, Eq.~(14))
\be\label{25}
\frac{d}{dr}\left(\mathcal{P} \, \frac{d\zeta}{dr}\right)+
            \left(\mathcal{Q}+\omega^2 \, \mathcal{W}\right)\, \zeta=0,
\ee
which is a second-order ordinary differential equation in the function $\zeta(r)$. 
The functions $\mathcal{P}, \mathcal{Q}$, and $\mathcal{W}$ are defined by (cf. \citep{KORU01}, Eqs.~(16), (17), and (15), respectively)
\begin{alignat}{1}
\label{26}
\mathcal{P} &= e^{(\lambda+3\nu)/2}  \,\, \frac{\Gamma \, P}{r^2},\\
\label{27}
\mathcal{Q} &= -e^{(3\nu+\lambda)/2} \,\, \frac{4}{r^3} \, \frac{dP}{dr}-e^{3(\nu+\lambda)/2} \,\, \frac{8 \pi}{r^2} \, P\, (E+P) + 
\frac{e^{(3\nu+\lambda)/2}}{r^2 \, (E+P)} \left( \frac{dP}{dr} \right)^2, \\
\label{28}
\mathcal{W}&=e^{(3\lambda+\nu)/2} \,\, \frac{E+P}{r^2}.
\end{alignat}
The adiabatic index $\Gamma$ involved in Eq.~(\ref{26}) is equal to 
\be\label{29}
\Gamma=\frac{E+P}{P} \,\, \frac{d\mathfrak{P}}{dE}
\ee
due to Eq.~(\ref{22}).

In order for the solution to be regular at the origin, the function $\zeta$ must obey the boundary conditions (\citep{KORU01}, paragraph following Eq.~(17))
\be
\zeta(r = 0) = 0, \qquad \qquad \zeta'(r = 0) = 0.
\label{SLIC}
\ee
At the surface of the star the Lagrangian variation of the pressure should vanish, $\Delta P(r=R)=0$, leading in turn to the condition (\citep{KORU01}, Eq.~(13a))
\be
\Gamma \, P \, \frac{d}{dr}\zeta(R) = 0.
\label{SLBC}
\ee

The discrete values $\omega_n^2$, $n=0,\,1,\,2,\,\dots$, for which the conditions~\eqref{SLIC} and \eqref{SLBC} are satisfied, form the set of eigenvalues  (eigenfrequencies) of the Sturm-Liouville problem established on Eqs.~\eqref{25}, \eqref{SLIC}, and \eqref{SLBC}. The corresponding eigenfunctions, $\zeta_n(r)$, have $n$ nodes inside the star and they are orthogonal. If $\omega^2>0$, then $\omega$ is real and the solution is purely oscillatory. However, if $\omega^2<0$, then $\omega$ is imaginary and corresponds to an exponentially growing solution. In this case we have an unstable neutron star.

\subsection{The adiabatic index $\Gamma$}
As discussed in \citep{CHN77} (Sec.~III), for different physical conditions inside a star, different adiabatic indices can be defined; provided, however, that the configuration has adequate time to attain equilibrium during the perturbation, the adiabatic index $\Gamma$ is related to an EOS $\mathfrak{P}(E)$ via Eq. \eqref{29}. This expression seems to be correct only for sufficiently low-frequency oscillations (see e.g. \cite{GLLI83}, Sec.~I, and also \cite{METH66}, Sec.~IIb). 

For central densities in the neutron drip density region, $10^{11}-10^{13}\,\mathrm{g \, cm^{-3}}$, the adiabatic index varies considerably \citep{GLLI83}. On the other hand, for central densities $\geq 10^{13} \, \mathrm{g \, cm^{-3}}$ the adiabatic index starts converging to the form~(\ref{29}).
Consequently, the relation~\eqref{29} seems to be suitable for the case of neutron stars when $\Gamma$ is to be treated as a variable.

\section{The Numerical Treatment}
The numerical method used in this study proceeds with two steps. In the first step, the nonrotating model is computed. In the second step, the radial oscillations are computed. Due to the purpose of the present investigation, we use two discrete forms for the polytropic EOS, as discussed in Sec~\ref{polEOS}; namely,
\begin{enumerate}
\item the polytropic EOS when the mass-energy density $E$ is involved (Eq.~\eqref{19});
\item the polytropic EOS when the rest-mass density $\rho$ is involved (Eq.~\eqref{17}).
\end{enumerate}

\subsection{Computing  the nonrotating models}
To compute the nonrotating models, we follow the numerical treatment regarding Hartle's perturbation method \citep{H67,HARTLEII} described in detail and used in the computations of  \citep{GK08}. We employ the parts of the theory and the computations concerning the nonrotating neutron star models (\citep{GK08}, Sec.~2 for the theoretical preliminaries and Sec.~5.1 for the particular computations). We do not intend to repeat here details on this issue.

\subsection{Computing the radial pulsation eigenfrequencies}
\label{ro}
To compute the eigenvalues $\omega^2$, we work as follows. We start the numerical integration for a trial value  $\omega^2$ and initial conditions~\eqref{SLIC}. We integrate towards the surface and then check if the resulting solution $\zeta(r)$ satisfies the boundary condition $\Gamma \, P \, \zeta(R)' = 0$ (Eq.~\eqref{SLBC}). From the point of view of numerical analysis, this boundary condition can be treated as an algebraic equation of the form $f\left( \omega^2 \right) = 0$; thus, to compute the root(s) $\omega^2$ of this equation, we can use a standard numerical method.
Before we apply such a method, we transform Eq.~\eqref{25} into a system of two first-order differential equations. This is achieved by introducing a new variable $\eta$ defined as
\be\label{41}
\eta = \mathcal{P} \, \frac{d\zeta}{dr}.
\ee
We thus obtain the differential equations (\citep{KORU01}, Eqs.~(19) and (20), respectively)
\be\label{42}
\frac{d\zeta}{dr} = \frac{\eta}{\mathcal{P}},
\ee
\be\label{43}
\frac{d\eta}{dr} = - \left( \omega^2 \, \mathcal{W} + \mathcal{Q} \right) \zeta.
\ee

Furthermore, we can transform the boundary condition \eqref{SLBC} into a form appropriate for the system of Eqs.~\eqref{42} and \eqref{43}. In particular, substituting Eq.~\eqref{41} into Eq. \eqref{SLBC}, and taking into account Eq.~\eqref{26}, we get the boundary condition
\be\label{44}
\left[ 
\frac{r^2}{e^{\left(\lambda+3\nu\right)/2}}
\right]
\eta(R)=0.
\ee

Next, through Taylor expansions, we find that the functions $\zeta$ and $\eta$ must have the following behavior near the origin (\citep{KORU01}, discussion following Eq.~(20)): $\zeta(r)=\zeta_0r^3+\mathcal{O}(r^5)$ and $\eta(r)=\eta_0+\mathcal{O}(r^2)$. Using Eq.~\eqref{42}, we find for the leading coefficients the relation $3 \, \zeta_0 = \eta_0 / \mathcal{P}(0)$.
Choosing 
\be
\eta_0 = 1, 
\label{h0}
\ee
we obtain 
\be
\zeta_0 = \frac{1}{3 \, \mathcal{P}\left( 0 \right)}.
\label{z0}
\ee

\subsection{The overall numerical method}
\label{onm}
 The overall method proceeds by integrating first the ``initial value problem'' (IVP, IVPs) established on the equations~\eqref{5}--\eqref{7}, using the numerical method described in detail in \citep{GK08} (Sec.~5.1). Second, for a trial value of $\omega^2$, we solve the IVP established on the equations~\eqref{42}--\eqref{43} with initial conditions~\eqref{h0}--\eqref{z0}.
Alternatively, we can solve all differential equations together, i.e. we can solve a unique IVP established on the five first-order differential equations~\eqref{5}--\eqref{7} and \eqref{42}--\eqref{43}.

The first procedure requires the functions $E(r)$, $m(r)$ and $\Phi(r)$ of the first IVP to be involved in the second IVP through their interpolating functions, which, apparently, increase the errors of the computations. To avoid interpolation errors,  we use in this study the alternative procedure according to which a unique IVP is solved for each trial value of $\omega^2$. We can then proceed to the numerical framework described in Sec.~\ref{ro}, i.e. to the rootfinding problem of the algebraic equation $f(\omega^2)=0$.

\subsection{The code}
%To compile our code, we have used the gfortran compiler, licensed under the GNU General Public License (GPL; http://www.gnu.org/licenses/gpl.html). gfortran is the name of the GNU Fortran compiler belonging to the GNU Compiler Collection (GCC; http://gcc.gnu. org/).  

Subroutines required for all numerical procedures of this study (e.g. solution of systems of first-order differential equations, interpolations of functions, rootfinding of algebraic equations, etc.) are taken from the SLATEC Common Mathematical Library, which is an extensive public-domain Fortran Source Code Library, incorporating several public-domain packages. The full SLATEC release is available in http://netlib.cs.utk.edu/.

\section{Numerical Results, Comparisons, and Discussion}
We compute the following models:
\begin{enumerate}
\item Model~I---The mass-energy density enters the polytropic EOS and the adiabatic index is treated as a variable,
\be P=KE^{1+\left(1/n\right)}\quad \text{and} \quad \Gamma=\frac{E+P}{P} \, \frac{dP}{dE}.
\ee
\item Model~II---The rest-mass density enters the polytropic EOS and the adiabatic index is treated as a variable,
\be
P=K\rho^{1+\left(1/n\right)}\quad \text{and} \quad \Gamma=\frac{E+P}{P} \, \frac{dP}{dE}.
\ee
\item Model~III---The mass-energy density enters the polytropic EOS and the adiabatic index is treated as a constant,
\be
P=KE^{1+\left(1/n\right)}\quad \text{and} \quad \Gamma=1+\frac{1}{n}.\ee
\item Model~IV---The rest-mass density enters the polytropic EOS and the adiabatic index is treated as a constant,
\be
P=K\rho^{1+\left(1/n\right)}\quad \text{and} \quad \Gamma=1+\frac{1}{n}.
\ee
\end{enumerate}

In the following tables, inputs occupied by ``--'' indicate negative eigenvalues, i.e. e-folding times, which are not examined in the present study.

For the purpose of testing our code, we compare results of Model I for $n=1.0$ with corresponding results given in \citep{KORU01} (Table A.18); and results of Model IV for $n=1.5$ with corresponding results given in \citep{HARTLEVIII} (Table 3). In all cases compared, the absolute percent differences, $\sigma$, of the results computed in the present study with respect to corresponding results of the other investigations, do not exceed the value $1.5\%$; in fact, there is only one case with $\sigma = 1.46\%$, while all other cases have $\sigma < 1.2\%$.

\begin{table}
\caption{Comparison of the computed eigenfrequencies for Model I,  when $n=1.0$, with respective results given by Kokkotas and Ruoff (2001, Table A.18; abbreviated ``KR''); $\sigma$ is the absolute percent difference of the values computed in the present study with respect to the values of KR.\label{ta9}}
\begin{center}
\begin{tabular}{lrrrr} 
\hline \hline 
$E_c$ (gu)    & 1.485$(-13)$  & 2.227$(-13)$    & 2.970$(-13)$   & 3.712$(-13)$   \\
$E_c$ (cgs) & 2.000(+15) & 3.000(+15)   & 4.000(+15)  & 5.000(+15)  \\
\hline 
$\nu_0$ (Hz)--Ref.~KR  & 2.323(+03) & 2.141(+03)  & 1.755(+03)  & 1.129(+03)  \\
$\nu_0$ (Hz)--present & 2.319(+03) & 2.136(+03)  & 1.748(+03)  & 1.118(+03)   \\
$\sigma$        & 0.17          & 0.23           & 0.40           & 0.97     \\
\hline 
$\nu_1$ (Hz)--Ref.~KR & 6.237(+03) & 6.871(+03)  & 7.244(+03)   & 7.475(+03)   \\
$\nu_1$ (Hz)--present          & 6.228(+03) & 6.862(+03)  & 7.234(+03)   & 7.465(+03)   \\
$\sigma$        & 0.14          & 0.13           & 0.14            & 0.13            \\ 
\hline 
$\nu_2$ (Hz)--Ref.~KR & 9.295(+03) & 10.319(+03) & 10.950(+03) & 11.365(+03) \\
$\nu_2$ (Hz)--present & 9.281(+03) & 10.030(+03) & 10.930(+03) & 11.350(+03) \\
$\sigma$        & 0.15          & 0.18            & 0.18             & 0.13            \\
\hline 
\end{tabular} 
\end{center}
\end{table}

\begin{table}
\caption{Comparison of the computed eigenvalues for Model IV, when $n=1.5$, with respective results given by Hartle and Friedman (1975, Table 3; abbreviated ``HF''); $\sigma$ is the absolute percent difference of the values computed in the present study with respect to the values of HF.\label{ta10}}
\begin{center}
\begin{tabular}{lrrrr} 
\hline \hline 
$E_c$ (gu)        & 7.424$(-14)$  & 2.346$(-13)$    & 7.424$(-13)$   & 2.346$(-12)$   \\
$E_c$ (cgs)       & 1.000(+15) & 3.160(+15)  & 1.000(+16)  & 3.160(+16)  \\
\hline 
$\omega_0^2$ (gu)--Ref.~HF & 5.870$(-14)$  & 4.900$(-14)$   & --                 & --  \\
$\omega_0^2$ (gu)--present & 5.856$(-14)$  & 4.844$(-14)$   & --                 & --  \\
$\sigma$                   & 0.23          & 1.14           & --                 & --  \\
\hline 
$\omega_1^2$ (gu)--Ref.~HF & 4.260$(-13)$  & 9.350$(-13)$  & 1.650$(-12)$     & 2.250$(-12)$ \\
$\omega_1^2$ (gu)--present & 4.272$(-13)$  & 9.354$(-13)$  & 1.656$(-12)$     & 2.313$(-12)$ \\
$\sigma$                   & 0.27        & 0.04        & 0.36           & 0.28     \\ 
\hline 
$\omega_2^2$ (gu)--Ref.~HF & 9.390$(-13)$  & 2.120$(-12)$  & 3.960$(-12)$     & 5.880$(-12)$ \\
$\omega_2^2$ (gu)--present & 9.384$(-13)$  & 2.122$(-12)$  & 3.967$(-12)$     & 5.967$(-12)$ \\
$\sigma$                   & 0.06          & 0.08          & 0.16       & 1.46         \\
\hline 
\end{tabular} 
\end{center}
\end{table}

Next, comparison of the results of Model IV in the place of Model I for $n=1.0$ with corresponding results of \citep{KORU01} (Table A.18) reveals differences larger than the previous ones, namely $\sim 2\%$ for $\nu_2$, $\sim 2.5\%$ for $\nu_1$, and $\sim 10\%$ for $\nu_0$. Likewise, comparison of the results of Model I in the place of Model IV for $n=1.5$ with corresponding results of \citep{HARTLEVIII} (Table 3) does also give differences larger than the previous ones, namely $\sim 5\%$ for $\nu_2$, $\sim 6\%$ for $\nu_1$, and $\sim 10\%$ for $\nu_0$. 

The above comparisons are made for models with central densities $E_\mathrm{c}$ ``below and relatively close'' (having the meaning: ``less than a specific quantity, but of same order of magnitude to this quantity'') to the ``maximum-mass densities'' of these models. Note that the total mass $M$ of a model, treated as a function of the central density $E_\mathrm{c}$, $M = M(E_\mathrm{c})$, obtains a maximum $M_\mathrm{max}$ for a specific value $E_\mathrm{c}^\mathrm{max}$; such a model is called ``maximum-mass model'', and the central density yielding this model is called ``maximum-mass density''. It is suitable to express maximum-mass densities and maximum masses in pu, since, this way, they are independent of the specific values chosen for the polytropic constant $K$. To facilitate comparisons, we compute, by using a method described in \citep{GESF} (Sec.~4), the values of $E_\mathrm{c}^\mathrm{max}$ and $M_\mathrm{max}$ for Models I \& III, and II \& IV, respectively, and quote these values in Tables \ref{ta1}, \ref{ta3}, \ref{ta5}, and \ref{ta7}. 

Extending to the cases $n=2.0$ and 2.5 our comparisons of respective results of Models I and IV, we find that differences for $\nu_2$ remain near $\sim 3\%$, differences for $\nu_1$ remain near $\sim 3.5\%$, while differences for $\nu_0$ remain near $\sim 10\%$ for $n=2.0$ (as in the previous cases), but for $n=2.5$ there is an increase in the difference to $\sim 20\%$. 

It seems therefore that, regarding Models I and IV for $E_\mathrm{c}$ below and relatively close to $E_\mathrm{c}^\mathrm{max}$ --- cases being, in fact, the more interesting ones when considering neutron stars ---  the behavior in the differences of $\nu_2$ and $\nu_1$ turns to be a systematic behavior. On the other hand, the differences for the lowest eigenfrequency $\nu_0$ seem to deviate according to how much close to a maximum-mass model is the case studied. Note that for a maximum-mass model the lowest eigenfrequency $\nu_0$ has to be zero; thus the closer to a maximum-mass model the case studied, the faster the convergence of $\nu_0$ to zero. However, maximum-mass densities do not coincide for Models I and IV and, accordingly, one case may be already close to the maximum-mass density, while the other case far enough yet.   

It is furthermore worth remarking that Models II and III yield extremum values among those examined here. In particular, Model II yields maximum values, while Model III yields minimum ones, at least for the eigenfrequencies $\nu_0$ and $\nu_1$. The interesting point is that the behavior in the differences of respective results turns again to be a systematic behavior for the eigenfrequencies $\nu_2$ and $\nu_1$: $\sim 8\%$ and $\sim 10\%$, respectively, for $n=1.0$; $\sim 0.5\%$ and $\sim 1\%$, respectively, for $n=1.5$; $\sim 0.5\%$ and $\sim 2\%$, respectively, for $n=2.0$; and $\sim 0.5\%$ and $\sim 1\%$, respectively, for $n=2.5$. In fact, there is a convergence of the respective values of $\nu_2$, and likewise of $\nu_1$, as the polytropic EOS becomes more and more soft. Note that, among the cases examined here, ``most stiff'' is the polytropic EOS with $n=1.0$, and ``most soft'' is that with $n=2.5$; thus siffness decreases as the polytropic index $n$ increases. On the other hand, differences in the lowest eigenfrequency $\nu_0$ seem to be again large: $\sim 35\%$ for $n=1.0$, $\sim 25\%$ for $n=1.5$, $\sim 35\%$ for $n=2.0$, and $\sim 45\%$ for $n=2.5$. The explanation for such discrepancies is similar to that given above: For a maximum-mass model the lowest eigenfrequency $\nu_0$ has to be zero; so, the closer to a maximum-mass model the case studied, the faster the convergence of $\nu_0$ to zero. However, maximum-mass densities do not coincide for Models II and III, and, accordingly, one case may be close to the maximum-mass density, while the other case far enough.

Finally, comparing respective results between Models I and III on the one side, and between Models II and IV on the other side, we arrive at conclusions similar to those of the previous comparisons.     

As a summary, regarding the eigenfrequencies $\nu_2$ and $\nu_1$, we have verified that, when the specific values of $E_\mathrm{c}$ are below and relatively close to $E_\mathrm{c}^\mathrm{max}$ (which are the more interesting cases when considering neutron stars), adopting any one of the Models I--IV leads to compatible results, i.e. results diverging each other $\sim 3\%$ on the average, and $\sim 10\%$ at worst. On the other hand, however, determining the lowest eigenfrequency $\nu_0$ seems to depend strongly on which of Models I--IV is involved in this computation.

\begin{table}
\caption{Numerical results for physical characteristics of the four nonrotating polytropic models for $n=1.0$ and $K = 6.6732 \times 10^4 \, \mathrm{cgs} = 10^{12} \, \mathrm{gu}$. Models I, III: $E_\mathrm{c}^\mathrm{max} = 4.2048 \times 10^{-1} \, \mathrm{pu}$, $M_\mathrm{max} = 1.9952 \times 10^{-1} \, \mathrm{pu}$. Models II, IV: $E_\mathrm{c}^\mathrm{max} = 4.1958 \times 10^{-1} \, \mathrm{pu}$, $M_\mathrm{max} = 1.6373 \times 10^{-1} \, \mathrm{pu}$. Upper four rows: Models I, III. Lower four rows: Models II, IV.\label{ta1}}
\begin{center}
{\scriptsize
\begin{tabular}{cccccccc} 
\hline \hline 
               \multicolumn{2}{c}{$E_c$}& \multicolumn{2}{c}{$\rho_c$}&\multicolumn{2}{c}{$R$} &\multicolumn{2}{c}{$M$}  \\

               gu & pu & gu  & pu & gu & pu  & gu & pu \\
\hline 
               1.485(-13) & 1.485(-01) & & & 9.673(+05) & 9.673(-01) & 1.662(+05) & 1.662(-01) \\
  2.227(-13) & 2.227(-01) & & & 8.862(+05) & 8.862(-01) & 1.870(+05) & 1.870(-01) \\
  2.970(-13) & 2.970(-01) & & & 8.256(+05) & 8.256(-01) & 1.959(+05) & 1.959(-01) \\
  3.712(-13) & 3.712(-01) & & & 7.787(+05) & 7.787(-01) & 1.991(+05) & 1.991(-01) \\
\hline 
  1.705(-13) & 1.705(-01) & 1.485(-13) & 1.485(-01) & 9.288(+05) & 9.288(-01) & 1.468(+05) & 1.468(-01) \\
  2.723(-13) & 2.723(-01) & 2.227(-13) & 2.227(-01) & 8.423(+05) & 8.423(-01) & 1.600(+05) & 1.600(-01) \\
  3.852(-13) & 3.852(-01) & 2.970(-13) & 2.970(-01) & 7.787(+05) & 7.787(-01) & 1.636(+05) & 1.636(-01) \\
  5.090(-13) & 5.090(-01) & 3.712(-13) & 3.712(-01) & 7.301(+05) & 7.301(-01) & 1.631(+05) & 1.631(-01) \\
\hline 
\end{tabular}} 
\end{center}
\end{table}

\begin{table}
\caption{Numerical results for the eigenvalues of the lowest three modes for $n=1.0$, measured in gravitational units unless stated otherwise. Upper, second, third, and lower four rows: Models I, II, III, and IV, respectively.\label{ta2}}
\begin{center}
\begin{threeparttable}
{\scriptsize
\begin{tabular}{cccccccc} 
\hline \hline 
                & & \multicolumn{2}{c}{Mode 0}& \multicolumn{2}{c}{Mode 1} &\multicolumn{2}{c}{Mode 2} \\
    $E_c$       &  $\rho_c$  & $\omega_0$ & $\nu_0$\tnote{*} & $\omega_1$ & $\nu_1$\tnote{*} & $\omega_2$  & $ \nu_2$\tnote{*} \\
\hline 
1.485(-13) & & 4.860(-07) & 2.319(+03) & 1.305(-06) & 6.228(+03) & 1.945(-06) & 9.281(+03) \\
2.227(-13) & & 4.477(-07) & 2.136(+03) & 1.438(-06) & 6.862(+03) & 2.159(-06) & 1.030(+04)  \\
2.970(-13) & & 3.664(-07) & 1.748(+03) & 1.516(-06) & 7.234(+03) & 2.292(-06) & 1.093(+04)  \\
3.712(-13) & & 2.344(-07) & 1.118(+03) & 1.565(-06) & 7.465(+03) & 2.378(-06) & 1.135(+04)  \\
\hline   
1.705(-13) & 1.485(-13) & 6.118(-07) & 2.919(+03) & 1.386(-06) & 6.612(+03) & 2.041(-06) & 9.738(+03)  \\
2.723(-13) & 2.227(-13) & 6.739(-07) & 3.216(+03) & 1.575(-06) & 7.516(+03) &2.327(-06) & 1.110(+04)  \\
3.852(-13) & 2.970(-13) & 7.148(-07) & 3.410(+03) & 1.711(-06) & 8.165(+03) & 2.531(-06) & 1.208(+04)  \\
5.090(-13) & 3.712(-13) & 7.454(-07) & 3.556(+03) & 1.817(-06) & 8.668(+03) & 2.693(-06) & 1.285(+04)  \\ 
\hline 
1.485(-13)&  & 3.867(-07) & 1.845(+03) & 1.247(-06) & 5.952(+03) & 1.878(-06) & 8.959(+03) \\
2.227(-13)&  & 2.328(-07) & 1.111(+03) & 1.346(-06) & 6.422(+03) & 2.052(-06) & 9.788(+03)  \\
2.970(-13)&  & 1.390(-06) & 6.633(+03) & 2.144(-06) & 1.023(+04) & 2.843(-06) & 1.356(+04)  \\
3.712(-13)&  & 1.406(-06) & 6.709(+03) & 2.193(-06) & 1.046(+04) & 2.917(-06) & 1.392(+04)  \\
\hline   
1.705(-13) & 1.485(-13) & 4.374(-07) & 2.087(+03) & 1.274(-06) & 6.078(+03) & 1.912(-06) & 9.123(+03)  \\
2.723(-13) & 2.227(-13) & 3.568(-07) & 1.702(+03) & 1.399(-06) & 6.674(+03) & 2.123(-06) & 1.013(+04)  \\
3.852(-13) & 2.970(-13) & 1.749(-07) & 8.343(+02) & 1.472(-06) & 7.026(+03) & 2.257(-06) & 1.077(+04)  \\
5.090(-13) & 3.712(-13) & --                 & --                  & 1.518(-06) & 7.245(+03) & 2.350(-06) & 1.121(+04)  \\ 
\hline 
\end{tabular}} 
\begin{tablenotes}
\item[*] measured in Hz
\end{tablenotes}
\end{threeparttable}
\end{center}
\end{table}

\begin{table}
\caption{Numerical results for physical characteristics of the four nonrotating polytropic models for $n=1.5$ and $K = 5.3802 \times 10^9 \, \mathrm{cgs} = 3.3887 \times 10^{7} \, \mathrm{gu}$. Models I, III: $E_\mathrm{c}^\mathrm{max} = 8.8846 \times 10^{-2} \, \mathrm{pu}$, $M_\mathrm{max} = 3.2027 \times 10^{-1} \, \mathrm{pu}$. Models II, IV: $E_\mathrm{c}^\mathrm{max} = 7.1671 \times 10^{-2} \, \mathrm{pu}$, $M_\mathrm{max} = 2.6438 \times 10^{-1} \, \mathrm{pu}$. Upper four rows: Models I, III. Lower four rows: Models II, IV.\label{ta3}}
\begin{center}
{\scriptsize
\begin{tabular}{cccccccc} 
\hline \hline 
               \multicolumn{2}{c}{$E_c$}& \multicolumn{2}{c}{$\rho_c$}&\multicolumn{2}{c}{$R$} &\multicolumn{2}{c}{$M$}  \\

               gu&pu&gu&pu&gu&pu&gu&pu\\
\hline 
6.842(-14) & 1.350(-02) & & & 1.322(+06) & 2.977(+00) & 1.079(+05) & 2.428(-01) \\
1.999(-13) & 3.945(-02) & & & 1.009(+06) & 2.272(+00) & 1.349(+05) & 3.037(-01) \\
5.530(-13) & 1.091(-01) & & & 7.540(+05) & 1.698(+00) & 1.418(+05) & 3.192(-01) \\
2.146(-12) & 4.235(-01) & & & 5.113(+05) & 1.151(+00) & 1.205(+05) & 2.712(-01) \\
\hline 
7.424(-14) & 1.465(-02) & 6.842(-14) & 1.350(-02) & 1.293(+06) & 2.912(+00) & 9.841(+04) & 2.216(-01) \\
2.346(-13) & 4.629(-02) & 1.999(-13) & 3.943(-02) & 9.767(+05) & 2.199(+00) & 1.159(+05) & 2.609(-01) \\
7.424(-13) & 1.465(-01) & 5.530(-13) & 1.091(-01) & 7.261(+05) & 1.635(+00) & 1.137(+05) & 2.559(-01) \\
2.346(-12) & 4.629(-01) & 2.146(-12) & 4.234(-01) & 5.057(+05) & 1.139(+00) & 8.778(+04) & 1.976(-01) \\
\hline 
\end{tabular}} 
\end{center}
\end{table}

\begin{table}
\caption{Numerical results for the eigenvalues of the lowest three modes for $n=1.5$, measured in gravitational units unless stated otherwise. Upper, second, third, and lower four rows: Models I, II, III, and IV, respectively.\label{ta4}}
\begin{center}
\begin{threeparttable}
{\scriptsize
\begin{tabular}{cccccccc} 
\hline \hline 
                  &  & \multicolumn{2}{c}{Mode 0}& \multicolumn{2}{c}{Mode 1} &\multicolumn{2}{c}{Mode 2} \\
 $E_c$        &  $\rho_c$   &$\omega_0$ & $\nu_0$\tnote{*} & $\omega_1$ & $\nu_1$\tnote{*} &$\omega_2$ &$ \nu_2$\tnote{*} \\
\hline 
7.424(-14) & & 2.688(-07) & 1.282(+03) & 6.944(-07) & 3.313(+03) &1.025(-06) & 4.890(+03) \\
2.346(-13) & & 2.920(-07) & 1.393(+03) & 1.069(-06) & 5.099(+03) &1.598(-06) & 7.623(+03) \\
7.424(-13) & & 1.481(-06) & 7.068(+03) & 2.266(-06) & 1.081(+04) & 2.990(-06)& 1.427(+04) \\
2.346(-12) & & 1.767(-06) & 8.430(+03) & 2.810(-06) & 1.341(+04) & --                & --                  \\
\hline   
7.424(-14) & 6.842(-14) & 3.116(-07) & 1.487(+03) & 6.872(-07) & 3.279(+03) & 1.003(-06) & 4.787(+03)  \\
2.346(-13) & 1.999(-13) & 4.773(-07) & 2.277(+03) &1.064(-06)  & 5.078(+03) & 1.556(-06) & 7.425(+03)  \\
7.424(-13) & 5.530(-13) & 6.897(-07) & 3.291(+03) &1.526(-06)  & 7.281(+03) & 2.235(-06) & 1.066(+04)  \\
2.346(-12) & 2.146(-12) & 1.114(-06) & 5.314(+03) & 2.167(-06) & 1.034(+04) & --                 & --  \\ 
\hline 
7.424(-14) & & 2.336(-07) & 1.115(+03) & 6.792(-07) & 3.240(+03) & 1.009(-06) & 4.815(+03) \\
2.346(-13) & & --                 & --                  & 1.020(-06) & 4.869(+03) & 1.548(-06) & 7.384(+03) \\
7.424(-13) & & --                 & --                  & 1.346(-06) & 6.422(+03) & 2.124(-06) & 1.014(+04) \\
3.712(-13) & & --                 & --                  & 1.421(-06) & 6.782(+03) & 2.474(-06) & 1.181(+04) \\
\hline   
7.424(-14) & 6.842(-14) & 2.420(-07) & 1.155(+03) & 6.536(-07) & 3.118(+03) & 9.687(-07) & 4.622(+03)  \\
2.346(-13) & 1.999(-13) & 2.201(-07) & 1.050(+03) & 9.672(-07) & 4.615(+03) & 1.457(-06) & 6.950(+03)  \\
7.424(-13) & 5.530(-13) & --                 & --                  & 1.287(-06) & 6.141(+03) & 1.992(-06) & 9.503(+03)  \\
2.346(-12) & 2.146(-12) & --                 & --                  & 1.521(-06) & 7.256(+03) & 2.443(-06) & 1.166(+04) \\ 
\hline 
\end{tabular}} 
\begin{tablenotes}
\item[*] measured in Hz
\end{tablenotes}
\end{threeparttable}
\end{center}
\end{table}

\begin{table}
\caption{Numerical results for physical characteristics of the four nonrotating polytropic models for $n=2.0$ and $K = 10^{12} \, \mathrm{cgs} = 1.2913 \times 10^{5} \, \mathrm{gu}$. Models I, III: $E_\mathrm{c}^\mathrm{max} = 9.4615 \times 10^{-3} \, \mathrm{pu}$, $M_\mathrm{max} = 6.0205 \times 10^{-1} \, \mathrm{pu}$. Models II, IV: $E_\mathrm{c}^\mathrm{max} = 5.7648 \times 10^{-3} \, \mathrm{pu}$, $M_\mathrm{max} = 5.1548 \times 10^{-1} \, \mathrm{pu}$. Upper four rows: Models I, III. Lower four rows: Models II, IV.\label{ta5}}
\begin{center}
{\scriptsize
\begin{tabular}{cccccccc} 
\hline \hline 
               \multicolumn{2}{c}{$E_c$}& \multicolumn{2}{c}{$\rho_c$}&\multicolumn{2}{c}{$R$} &\multicolumn{2}{c}{$M$}  \\

               gu&pu&gu&pu&gu&pu&gu&pu\\
\hline 
7.424(-14) & 1.235(-03) & & & 1.366(+06) & 1.059(+01) & 6.689(+04) & 5.185(-01) \\
1.485(-13) & 2.471(-03) & & & 1.121(+06) & 8.690(+00) & 7.256(+04) & 5.625(-01) \\
2.970(-13) & 4.942(-03) & & & 9.138(+05) & 7.084(+00) & 7.643(+04) & 5.925(-01) \\
7.424(-13) & 1.235(-02) & & & 6.908(+05) & 5.355(+00) & 7.749(+04) & 6.007(-01) \\
\hline 
7.947(-14) & 1.322(-03) & 7.424(-14) & 1.235(-03) & 1.351(+06) & 1.047(+01) & 6.169(+04) & 4.782(-01) \\
1.633(-13) & 2.717(-03) & 1.485(-13) & 2.471(-03) & 1.106(+06) & 8.574(+00) & 6.520(+04) & 5.054(-01) \\
3.388(-13) & 5.638(-03) & 2.970(-13) & 4.942(-03) & 9.015(+05) & 6.988(+00) & 6.655(+04) & 5.159(-01) \\
9.076(-13) & 1.510(-02) & 7.424(-13) & 1.235(-02) & 6.842(+05) & 5.304(+00) & 6.426(+04) & 4.981(-01) \\
\hline 
\end{tabular}} 
\end{center}
\end{table}

\begin{table}
\caption{Numerical results for the eigenvalues of the lowest three modes for $n=2.0$, measured in gravitational units unless stated otherwise. Upper, second, third, and lower four rows: Models I, II, III, and IV, respectively.\label{ta6}}
\begin{center}
\begin{threeparttable}
{\scriptsize
\begin{tabular}{cccccccc} 
\hline \hline 
                  &  & \multicolumn{2}{c}{Mode 0}& \multicolumn{2}{c}{Mode 1} &\multicolumn{2}{c}{Mode 2} \\
  $E_c$        &  $\rho_c$   &$\omega_0$ & $\nu_0$\tnote{*} & $\omega_1$ & $\nu_1$\tnote{*} &$\omega_2$ &$ \nu_2$\tnote{*} \\
\hline 
7.424(-14) & & 1.725(-07) & 8.232(+02) & 4.875(-07) & 2.326(+03) & 7.145(-07) & 3.409(+03) \\
1.485(-13) & & 2.057(-07) & 9.817(+02) & 6.614(-07) & 3.156(+03) & 9.734(-07) & 4.645(+03)  \\
2.970(-13) & & 2.077(-07) & 9.910(+02) & 8.841(-07) & 4.218(+03) & 1.309(-06) & 6.244(+03)  \\
7.424(-13) & & 1.259(-07) & 6.005(+03) & 1.882(-06) & 8.982(+03) & 2.465(-06) & 1.176(+04)  \\
\hline   
7.947(-14) & 7.424(-14) & 2.194(-07) & 1.047(+03) & 4.921(-07) & 2.348(+03) & 7.126(-07) & 3.400(+03)  \\
1.633(-13) & 1.485(-13) & 3.024(-07) & 1.443(+03) & 6.705(-07) & 3.199(+03) & 9.705(-07) & 4.631(+03)  \\
3.388(-13) & 2.970(-13) & 4.139(-07) & 1.975(+03) & 9.020(-07) & 4.304(+03) & 1.304(-06) & 6.224(+03)  \\
9.076(-13) & 7.424(-13) & 6.206(-07) & 2.961(+03) & 1.300(-06) & 6.201(+03) & 1.876(-06) & 8.953(+03)  \\ 
\hline 
7.424(-14) & & 1.445(-07) & 6.896(+02) & 4.801(-07) & 2.291(+03) & 7.075(-07) & 3.376(+03) \\
1.485(-13) & & 1.371(-07) & 6.541(+02) & 6.474(-07) & 3.089(+03) & 9.601(-07) & 4.581(+03) \\
2.970(-13) & & 8.581(-07) & 4.094(+03) & 1.284(-06) & 6.126(+03) & 1.682(-06) & 8.024(+03) \\
7.424(-13) & & 1.202(-06) & 5.734(+03) & 1.828(-06) & 8.722(+03) & 2.408(-06) & 1.149(+04) \\
\hline   
7.947(-14) & 7.424(-14) & 1.509(-07) & 7.199(+02) & 4.704(-07) & 2.244(+03) & 6.924(-07) & 3.304(+03) \\
1.633(-13) & 1.485(-13) & 1.574(-07) & 7.511(+02) & 6.299(-07) & 3.006(+03) & 9.328(-07) & 4.451(+03) \\
3.388(-13) & 2.970(-13) & 2.035(-08) & 9.708(+01) & 8.276(-07) & 3.949(+03) & 1.236(-06) & 5.896(+03) \\
9.076(-13) & 7.424(-13) & --                 & --                  & 1.141(-06) & 5.444(+03) & 1.730(-06) & 8.255(+03) \\ 
\hline 
\end{tabular}} 
\begin{tablenotes}
\item[*] measured in Hz
\end{tablenotes}
\end{threeparttable}
\end{center}
\end{table}

\begin{table}
\caption{Numerical results for physical characteristics of the four nonrotating polytropic models for $n=2.5$ and $K = 1.5 \times 10^{13} \, \mathrm{cgs} = 2.9797 \times 10^{3} \, \mathrm{gu}$. Models I, III: $E_\mathrm{c}^\mathrm{max} = 2.9028 \times 10^{-4} \, \mathrm{pu}$, $M_\mathrm{max} = 1.3623 \, \mathrm{pu}$. Models II, IV: $E_\mathrm{c}^\mathrm{max} = 1.2553 \times 10^{-4} \, \mathrm{pu}$, $M_\mathrm{max} = 1.2442 \, \mathrm{pu}$. Upper four rows: Models I, III. Lower four rows: Models II, IV.\label{ta7}}
\begin{center}
{\scriptsize
\begin{tabular}{cccccccc} 
\hline \hline 
               \multicolumn{2}{c}{$E_c$}& \multicolumn{2}{c}{$\rho_c$}&\multicolumn{2}{c}{$R$} &\multicolumn{2}{c}{$M$}  \\

               gu& pu &gu&pu&gu&pu&gu&pu\\
\hline 
7.424(-14) & 3.599(-05) & & & 1.303(+06) & 5.918(+01) & 2.824(+04) & 1.283(+00) \\
1.485(-13) & 7.199(-05) & & & 1.049(+06) & 4.764(+01) & 2.915(+04) & 1.324(+00) \\
2.970(-13) & 1.440(-04) & & & 8.438(+05) & 3.832(+01) & 2.988(+04) & 1.357(+00) \\
7.424(-13) & 3.599(-04) & & & 6.304(+05) & 2.863(+01) & 2.997(+04) & 1.361(+00) \\
\hline 
7.734(-14) & 3.749(-05) & 7.424(-14) & 3.599(-05) & 1.302(+06) & 5.913(+01) & 2.682(+04) & 1.218(+00) \\
1.567(-13) & 7.596(-05) & 1.485(-13) & 7.199(-05) & 1.049(+06) & 4.764(+01) & 2.729(+04) & 1.239(+00) \\
3.185(-13) & 1.544(-04) & 2.970(-13) & 1.440(-04) & 8.450(+05) & 3.838(+01) & 2.738(+04) & 1.244(+00) \\
8.202(-13) & 3.976(-04) & 7.424(-13) & 3.599(-04) & 6.338(+05) & 2.879(+01) & 2.677(+04) & 1.216(+00) \\
\hline 
\end{tabular}} 
\end{center}
\end{table}

\begin{table}
\caption{Numerical results for the eigenvalues of the lowest three modes for $n=2.5$, measured in gravitational units unless stated otherwise. Upper, second, third, and lower four rows: Models I, II, III, and IV, respectively.\label{ta8}}
\begin{center}
\begin{threeparttable}
{\scriptsize
\begin{tabular}{cccccccc} 
\hline \hline 
               &  & \multicolumn{2}{c}{Mode 0}& \multicolumn{2}{c}{Mode 1} &\multicolumn{2}{c}{Mode 2} \\
    $E_c$        &  $\rho_c$   &$\omega_0$ & $\nu_0$\tnote{*} & $\omega_1$ & $\nu_1$\tnote{*} &$\omega_2$ &  $ \nu_2$\tnote{*} \\
\hline 
7.424(-14) & & 9.617(-08) & 4.589(+02) & 3.373(-07) & 1.607(+03) & 4.891(-07) & 2.334(+03) \\
1.485(-13) & & 1.160(-08) & 5.534(+02) & 4.676(-07) & 2.231(+03) & 6.794(-07) & 3.242(+03)  \\
2.970(-13) & & 1.198(-07) & 5.715(+02) & 6.442(-07) & 3.074(+03) & 2.968(-06) & 4.478(+03)  \\
7.424(-13) & & 9.719(-07) & 4.637(+03) & 1.423(-06) & 6.788(+03) & 1.847(-06) & 8.813(+03)  \\
\hline   
7.734(-14) & 7.424(-14) & 1.350(-07) & 6.442(+02) & 3.365(-07) & 1.606(+03) & 4.834(-07) & 2.306(+03)  \\
1.567(-13) & 1.485(-13) & 1.911(-07) & 9.120(+02) & 4.661(-07) & 2.222(+03) & 6.690(-07) & 3.192(+03)  \\
3.186(-13) & 2.970(-13) & 2.705(-07) & 1.291(+03) & 6.414(-07) & 3.061(+03) & 9.196(-07) & 4.388(+03)  \\
8.202(-13) & 7.424(-12) & 4.280(-07) & 2.042(+03) & 9.657(-07) & 4.608(+03) & 1.383(-06) & 6.592(+03)  \\ 
\hline 
7.424(-14) & & 7.434(-08) & 3.547(+02) & 3.344(-07) & 1.596(+03) & 4.865(-07) & 2.321(+03) \\
1.485(-13) & & 6.166(-08) & 2.942(+02) & 4.623(-07) & 2.206(+03) & 6.747(-07) & 3.219(+03)  \\
2.970(-13) & & 6.347(-07) & 3.028(+03) & 9.300(-07) & 4.437(+03) & 1.208(-06) & 5.764(+03)  \\
7.424(-13) & & 9.516(-07) & 4.540(+03) & 1.404(-06) & 6.700(+04) & 1.829(-06) & 8.727(+03)  \\
\hline   
7.734(-14) & 7.424(-14) & 7.687(-08) & 3.668(+02) & 3.263(-07) & 1.557(+03) & 4.745(-07) & 2.264(+03)  \\
1.567(-13) & 1.485(-13) & 7.114(-08) & 3.394(+02) & 4.477(-07) & 2.136(+03) & 6.530(-07) & 3.116(+03)  \\
3.186(-13) & 2.970(-13) & 6.085(-07) & 2.904(+03) & 8.911(-07) & 4.252(+03) & 1.158(-06) & 5.524(+03)  \\
8.202(-13) & 7.424(-13) & --                 & --                 & 8.962(-07) & 4.276(+03) & 1.321(-06) & 6.305(+03)  \\ 
\hline 
\end{tabular}} 
\begin{tablenotes}
\item[*] measured in Hz
\end{tablenotes}
\end{threeparttable}
\end{center}
\end{table}

\end{document}